\documentclass[12pt]{article}
\pdfoutput=1
\usepackage{putex}
\usepackage{amsmath}
\usepackage{amssymb}
\usepackage{graphicx}
\usepackage{epstopdf}
\usepackage{enumerate}
\usepackage{cite}
\usepackage{tensor}
\usepackage{slashed}
\usepackage{feynmf}
\usepackage{hyperref}
\usepackage{bbold}
\usepackage{dsfont}
\usepackage{mathrsfs}
\usepackage{caption}
\usepackage{subcaption}

\usepackage{youngtab}
\usepackage{amsfonts}
\usepackage{braket}

\usepackage[utf8x]{inputenc}
\usepackage{braket}
\usepackage{titlesec}
\usepackage{fixltx2e}
\usepackage{cleveref}
\hypersetup{breaklinks}

\usepackage[usenames,dvipsnames,svgnames,table]{xcolor}

\def\p{\partial}

\newcommand {\be} {\begin {equation}}
\newcommand {\ee} {\end {equation}}
\newcommand{\bea}{\begin{eqnarray}}
\newcommand{\eea}{\end{eqnarray}}

\def\eps{\epsilon}
\def\zb{\overline{z}}

\def\rt{\rightarrow}

 \newmuskip\pFqmuskip

\newcommand*\pFq[6][8]{%
  \begingroup 
  \pFqmuskip=#1mu\relax
  \mathcode`\,=\string"8000
  \begingroup\lccode`\~=`\,
  \lowercase{\endgroup\let~}\pFqcomma
  {}_{#2}F_{#3}{\left[\genfrac..{0pt}{}{#4}{#5};#6\right]}%
  \endgroup
}
\newcommand{\pFqcomma}{\mskip\pFqmuskip}

\makeatletter
\renewcommand{\@maketitle}{
\newpage
 \begin{center}%
  {\large\bfseries \@title \par}%
 \end{center}%
 \par} \makeatother

\numberwithin{equation}{section}

\titleformat*{\section}{\large\bfseries}

\begin{document}

\institution{UCLA}{ \quad\quad\quad\quad\quad\quad\quad\quad\quad Mani L. Bhaumik Institute for Theoretical Physics, \cr Department of Physics and Astronomy, University of California, Los Angeles, CA 90095, USA}

\title{Anomalous dimensions from  \\ quantum Wilson lines}

\authors{Mert Besken, Ashwin Hegde and  Per Kraus}

\abstract{We study the self-energy of a gravitating point particle in AdS$_3$, and compare to operator dimensions in CFT$_2$.  In particular, we compute the one and two loop diagram contributions to the expectation value of an open Wilson line in the SL(2,R)$\times$ SL(2,R) Chern-Simons formulation of AdS$_3$ gravity.  This gives the two-point function of CFT primary operators to second order in a large $c$ expansion, and hence yields the scaling dimension $h(j,c)$ as a function of the SL(2,R) spin $j$.  Comparison to CFT is made in the context of constructing Virasoro representations starting from representations of SL(2,R) current algebra.  Our Wilson line computations follow the framework advanced recently by Fitzpatrick et. al., which is based on earlier work by H. Verlinde.  We encounter some renormalization scheme ambiguities at the two-loop level which we are not able to fully resolve, hampering a definitive comparison with CFT expressions at this order.        }

\date{}

\maketitle
\setcounter{tocdepth}{2}
\tableofcontents

\section{Introduction}

In this paper we study the gravitational self-energy of a point particle in AdS$_3$, and in particular the relation between the energy of the particle when Newton's constant is vanishing or finite. Typically, the relation between these energies is not very interesting since it is cutoff dependent: the self-energy suffers from the classic UV divergence problem, necessitating a short distance cutoff, and there is no universal relation between the bare and renormalized energies.  However, for a  particle in AdS$_3$ the situation appears to be more favorable, as we now discuss.

The Hilbert space of a particle coupled to gravity in AdS$_3$ corresponds, via the AdS$_3$/CFT$_2$ duality, to a representation of the Virasoro algebra.  The lowest allowed energy  of the particle maps to the dimension of the primary operator which labels the representation, $E_0=h+\overline{h}-{c\over 12}$.     We will use the well-known fact \cite{Bershadsky:1989mf}, reviewed below, that representations of the Virasoro algebra can be obtained by starting from SL(2,R) current algebra and imposing constraints on the currents.  Starting from an SL(2,R) primary of spin-$j$, one thereby obtains a Virasoro primary of dimension $h(j,c)$, which depends on $j$ and the central charge $c$.  The formula can be written as
\be h(j,c) = -j +{m+1\over m}j(j+1)~,\quad c=1-{6\over m(m+1)}~.\ee
Recalling the Brown-Henneaux formula \cite{Brown:1986nw}, $c=3l/2G_N$, sending $G_N\rt 0$ corresponds to  $c\rt \infty$, which can be accomplished by taking $m\rt -1$.  $h(j,c)$ admits an expansion in $1/c$,
\be\label{hres} h(j,c) = -j -{6\over c}j(j+1)-{78\over c^2} j(j+1)+\ldots~.
\ee
We aim to give the subleading terms  an interpretation in terms of gravitational self-energy.\footnote{In much of this paper we will take $2j$ to be a positive integer corresponding to a finite dimensional non-unitary representation of SL(2,R).  This of course yields a negative ``bare" energy.  However, we stress that our analysis carries over immediately to j-values corresponding to positive energy unitary representations, as we discuss later.  }

The relation between the SL(2,R) current algebra and the Virasoro algebra has an analog on the AdS$_3$ side that is also well known; see \cite{Banados:1998gg} for a review.  Starting from SL(2,R)$\times$ SL(2,R) Chern-Simons theory, which is equivalent \cite{Witten:1988hc,Achucarro:1987vz} (in perturbation theory) to three-dimensional Einstein gravity with a negative cosmological constant, imposing the boundary conditions that imply asymptotic AdS$_3$-ness has the effect of implementing the aforementioned reduction of the symmetry algebra.   Our particle is described by a Wilson line in the spin-$j$ representation of SL(2,R).   An open Wilson line with endpoints on the AdS boundary  computes a boundary two-point function, from which the dimension $h(j,c)$ can be deduced, and hence our task is to compute such a Wilson line perturbatively in $1/c$.  Wilson lines in the context of AdS$_3$/CFT$_2$ duality first appeared in \cite{Ammon:2013hba,deBoer:2013vca} as a tool to compute entanglement entropy in higher spin theories.

 Our setup is motivated by ongoing work  \cite{Hartman:2013mia,Fitzpatrick:2014vua,deBoer:2014sna,Hijano:2015rla, Fitzpatrick:2015zha,Alkalaev:2015wia,Hijano:2015qja,
Fitzpatrick:2015dlt,
Bhatta:2016hpz} on the bulk interpretation of conformal blocks in two-dimensional CFTs, which is in turn aimed at gaining insight into the emergence of local bulk physics --- and its ultimate breakdown --- starting from CFT.  In particular, conformal blocks were given a bulk formulation in terms of particle worldlines in \cite{Fitzpatrick:2014vua,Hijano:2015rla,Fitzpatrick:2015zha,Alkalaev:2015wia,
Hijano:2015zsa,Hijano:2015qja,Alkalaev:2015lca,Alkalaev:2016ptm,
Fitzpatrick:2016ive,Czech:2016xec,daCunha:2016crm,Chen:2016dfb,Alkalaev:2016rjl,Guica:2016pid}.  A Wilson line version of these constructions in the large $c$ limit, with generalizations to higher spin theories,  was given in \cite{deBoer:2014sna,Hegde:2015dqh,Bhatta:2016hpz,Besken:2016ooo}.  The fully quantum version incorporating $1/c$ corrections appears in \cite{Fitzpatrick:2016mtp}.  We should also note that the main features of these Wilson line constructions already appeared long ago in  \cite{Verlinde:1989ua}, building on the famous connection between Chern-Simons theory and CFT developed in \cite{Witten:1988hf}, albeit at a somewhat formal level that did not take into account such issues as UV divergences.  This early work is reviewed in the modern AdS/CFT context in \cite{Fitzpatrick:2016mtp}.
\begin{figure}[h]
  \flushleft
  \begin{minipage}[b]{0.68\textwidth}
    \centering\includegraphics[width=1.4\textwidth]{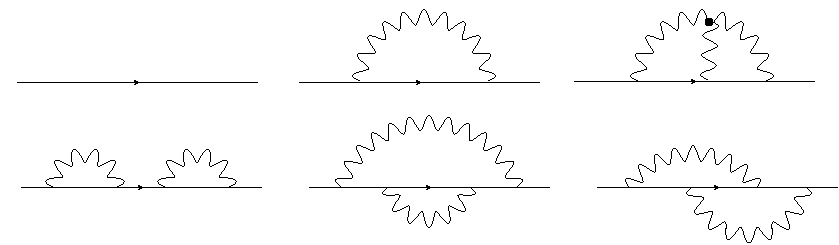}
  \end{minipage}
  \caption{ Wilson line diagrams to order $1/c^2$ }
  \label{fig:fa}
\end{figure}

\begin{figure}[h]
 \quad \quad \quad \quad \quad \quad
  \begin{minipage}[b]{0.5\textwidth}
    \centering\includegraphics[width=1.4\textwidth]{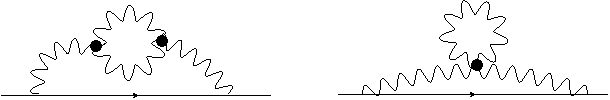}
  \end{minipage}
  \caption{ Graviton self energy }
  \label{fig:fb}
\end{figure}

We compute a Wilson line two-point function to the first two subleading  orders in the $1/c$ expansion, corresponding to the diagrams shown in figure \ref{fig:fa}.\footnote{The graviton self-energy diagrams in figure \ref{fig:fb} are implicitly taken into account, as will become clear.}
 These diagrams are UV divergent, as expected.  The proper treatment of these divergences is not completely straightforward, as we are not starting from the standard framework of a local Lagrangian to which we can add counterterms, and this gives rise to some ambiguities.  At order $1/c$ simply removing power law divergences yields the first correction in (\ref{hres}).  At order $1/c^2$ the two loop diagrams include contributions that can be unambiguously associated to the exponentiation of the order $1/c$ result, but ambiguity arises in trying to deduce the $1/c^2$ correction to $h(j,c)$, essentially due to the need to remove a divergent term of the same form as the finite term we are after. It seems likely that to resolve this ambiguity one needs to study in more detail how the Virasoro generators act in this setup and require that the symmetry is being implemented consistently.

\section{CFT results}

We begin by reviewing how imposing constraints on SL(2,R) current algebra representations yields representations of Virasoro \cite{Bershadsky:1989mf}.  The SL(2,R) current algebra at level $k$ is
\be J^a(z)J^b(0) \sim  {(k/2 )\eta^{ab} \over z^2}  + {i \eps^{ab}_{~~c}J^c(0)\over z} \ee
Here $\eta^{ab}=(1,1,-1)$ and $\eps^{123} =1$.   We also define $J^\pm = J^1 \pm i J^2$.
The stress tensor is given via the Sugawara construction
\be  T_{\rm SL(2)} =  {1\over k-2} \eta_{ab}J^a J^b \ee
Its modes obey a Virasoro algebra with central charge
\be  c_{\rm SL(2)} = {3k\over k-2} \ee
Current algebra primaries sit in representations of SL(2,R), as labelled by the quadratic Casimir $C_2 = \eta_{ab}J^a J^b$  and the $J^3$ eigenvalue.  For ease of comparison with our later formulas it turns out to be convenient to focus on representations with $J^3$ bounded from below, and  to define $j$ as the negative of the smallest value of $J^3$ in the representation, so that the quadratic Casimir is $C_2 = -j(j+1)$.  In this notation, taking $2j$ to be a positive integer yields a finite dimensional, non-unitary, representation of SL(2,R).  The scaling dimension of a spin-j primary is
\be  h_{\rm SL(2)}[\Phi_j] = - {j(j+1)\over k-2} \ee

The reduction to Virasoro proceeds by imposing the constraints $J^-(z)=k$ and $J^0(z)=0$.   For conformal invariance to be compatible with the $J^-$ constraint the stress tensor needs to be modified so that $J^-(z)$ acquires vanishing scaling dimension.  This is accomplished by adding to the stress tensor a term proportional to $\p J^3(z)$.  Also, ghosts are introduced so that the constraints can be implemented by a BRST construction. The full stress tensor is then
\be
T = T_{\rm SL(2)} + \p J^3 + T_{\rm gh} \ee
with central charge
\be  c = {3k\over k-2} + 6k -2  \ee
with the $-2$ coming from the ghosts.   The improvement term yields a contribution $J^3$ to the dimension of the original current algebra primaries, so the dimension of the Virasoro primary is
\be  h[\Phi_j] =-j - {j(j+1)\over k-2} \ee
since $J^3=-j$ yields the lowest dimension operator.    It is convenient to write the central charge in the standard minimal model parametrization
\be
c = 1- {6\over m (m+1)}~, \quad k= {m+2\over m+1}~,
\ee
so that
\be\label{hform}  h(j,c)\equiv h[\Phi_j] =-j+ {m+1 \over m}j(j+1) \ee

To put this in context, recall that the dimensions of the Kac degenerate representations are
\be
h_{r,s} = { \big(r(m+1)-sm\big)^2-1\over 4m (m+1)}~. \ee
We have
\be
h(j,c) = h_{r,s}~,\quad r=2j+1~,~ s=1~.
\ee

Of interest to us is the large $c$ limit obtained by taking $m\rt -1$, which yields
\be\label{hexp} h(j,c)] = -j - {6j(j+1)\over c} - {78 j(j+1)\over  c^2}+ \ldots~.
\ee
The alternative case $m\rt 0$ is commented on below.  As we have discussed, we expect the  terms appearing in the expansion (\ref{hexp}) to correspond, in the bulk, to perturbative gravitational self energy diagrams.

\section{Bulk side: preliminary comments}

The $1/c$ expansion on the CFT side maps to an expansion in the 3d Newton constant $G$, so we can hope to recover (\ref{hexp}) by gravitational perturbation theory in AdS$_3$.  The Brown-Henneaux formula  $c=3l/2G$ relates the expansions.\footnote{More precisely, we should recall that the Brown-Henneaux formula is a classical result in Einstein gravity.   In the presence of higher derivative terms it is replaced by the Wald-like formula \cite{Saida:1999ec,Kraus:2005vz}  $c= {l\over 2G} g_{\mu\nu}   {\delta {\cal L} \over \delta R_{\mu\nu}}$.}

Let us first give a heuristic explanation for the part of (\ref{hexp}) which is due to classical self-energy.  We consider a spinless point particle of mass $ml=2h\gg 1$.  In higher than three dimensions, as soon as gravity is turned on the particle would collapse into a black hole, but in three dimensions and for sufficiently light particles one instead gets a conical defect solution.   In the absence of a cosmological constant, a particle of mass $m$ yields a solution described by Minkowski space with a wedge of angle $\Delta \phi = 8\pi G m$ cut out \cite{Deser:1983tn}.  Let us now think of placing this particle in AdS$_3$.  We do so while keeping $m$ fixed, meaning that we hold fixed the deficit angle computed by examining the geometry in the immediate neighborhood of the solution.  Now, a conical defect solution in AdS$_3$ takes the form
\be ds^2= -(r^2-8GMl^2)dt^2+{l^2dr^2 \over r^2-8GMl^2} +r^2 d\phi^2 \ee
where $\phi \cong \phi+2\pi$.   Here $M$ is the total energy  measured at the asymptotic AdS boundary, with $M<0$ for a conical defect.  By rescaling coordinates, this metric can be written in standard form $ds^2 = -(r^2+l^2)dt^2 + l^2dr^2/(r^2+l^2) +r^2d\phi$ but with an angle $\Delta \phi = 2\pi(1-\sqrt{-8GM})$ cut out. Equating our two expressions for $\Delta \phi$ yields the relation between the ``bare" mass $m$ and the physical energy $M$,
\be M = -{1\over 8G} +m -2Gm^2~. \ee
Writing $Ml=-{c\over 12}+2h$, $ml=-2j$, and using the Brown-Henneaux formula, this becomes
\be h = -j-{6j^2\over c}~. \ee
The $j^2/c$ contribution matches (\ref{hexp}).  To capture the $j/c$ term we need to go beyond treating the particle as having a definite position and include the effect of its finite size quantum wavefunction, which is suppressed for $j\gg 1$.  This effect is incorporated in the perturbative treatment given below.

Before turning to that analysis let us return to (\ref{hform}) and now  expand around $m\rt 0$,
\be h(j,c) =  -{j(j+1) \over 6}c   -j+{13 j(j+1)\over 6} +{6j(j+1)\over c}+\ldots~.\ee
This result was given a nice bulk interpretation in \cite{Raeymaekers:2014kea}; to compare, set $j=(s-1)/2$ and write $L_0 = h(j,c) -{c\over 24} = -{s^2 c\over 24} +{(13s+1)(s-1)\over 24} + \ldots$. These states correspond to classical solutions with conical excess angle $2\pi(s-1)$.  The $O(c^0)$ contribution comes from quantizing the solutions using the method of coadjoint orbits.  For $s$ a positive integer these representations correspond to the degenerate $h_{1,s}$ representations of the Virasoro algebra, examined at large $c$.

\section{Perturbative self-energy computation}

\subsection{Chern-Simons formulation, and correlators from Wilson lines}

The Chern-Simons formulation of 3d gravity is perfectly adapted to our problem, since the above procedure of going from SL(2,R) current algebra to Virasoro has a precise counterpart in terms of imposing boundary conditions on the connection in SL(2,R)$\times $ SL(2,R) Chern-Simons theory.   In the bulk, the Virasoro symmetry arises as the symmetry algebra preserving the asymptotic boundary conditions.   We will not review the details of this, as it is well described in many references, e.g. \cite{Banados:1998gg}.   We just note the following.  AdS$_3$ in the form $ds^2 = d\rho^2 +e^{2\rho}dzd\zb$ is represented by the connection $A= L_0 d\rho + e^\rho L_1 dz$, along with a similar expression for the second SL(2,R) factor which we henceforth suppress.  Here $L_n$ are standard SL(2,R) generators obeying $[L_m,L_n]=(m-n)L_{m+n}$.   More generally, a metric with boundary stress tensor $T(z)$ is represented by $A = L_0 d\rho + (L_1+ {6\over c}T(z) e^{-\rho} L_{-1})dz$.  The $\rho $ dependence can be removed by a gauge transformation by $e^{ \ln(\rho)L_0}$, allowing us  to work with the reduced connection
\be\label{adef} a = \big(L_1+{6\over c}T(z)L_{-1}\big) dz~.\ee

Given a connection of the above form, the rule for computing correlators is extremely simple.  More precisely, we focus here on the conformal blocks, and in particular just the holomorphic half of the conformal block.   Each operator in the CFT  corresponds to some spin-j representation of SL(2,R).  

In the large $c$ limit the rule for computing conformal blocks is as follows \cite{Bhatta:2016hpz,Besken:2016ooo}.  We set $T(z)=0$ corresponding to the vacuum state.   Each primary operator is represented by its corresponding highest weight SL(2,R) state $|j_i j_i\rangle$.  We then attach a Wilson line $W_{j_i}[z_i,z_b]=P e^{\int_{z_i}^{z_b} a}$ directed from the operator location to some arbitrary location $z_b$.  At $z_b$ there resides a singlet state $\langle S|$   in the tensor product of the representations of the primary operators.  The large $c$ conformal block is then simply
\be\label{Bdef} G(z_1,j_1; z_2, j_2; \ldots z_n , j_n) = \langle S| \prod_{i=1}^n W_{j_i}[z_i,z_b] |j_i j_i\rangle~. \ee
This expression satisfies two basic properties. First,  it is independent of the choice of $z_b$, as moving $z_b$ is easily seen to be realized by a gauge transformation, which acts trivially on the singlet state. Second, gauge invariance implies that it transforms as it should under conformal transformations.  We also remark that there are in general multiple ways to construct singlet states out of the representations hosted by the primary operators, and this corresponds to the space of conformal blocks.  A full fledged correlation function is constructed by combining holomorphic and anti-holomorphic conformal blocks in a manner compatible with crossing symmetry.

The above large $c$ construction yields the global conformal blocks, in which exchanged operators fill out representations of the global conformal group SL(2,R).  These conformal blocks can be viewed as the large $c$ limit of Virasoro blocks, which are much richer objects.  From the bulk point of view, the Virasoro blocks capture the effect of gravitational interactions, including both classical and quantum effects.  Indeed, at finite $c$ the Virasoro blocks in some sense contain non-perturbative quantum gravity effects \cite{Fitzpatrick:2016mjq,Fitzpatrick:2016ive}, and indeed this is the main motivation for trying to formulate them in bulk terms.

At finite $c$ the same construction (\ref{Bdef}) applies, at least formally, except now we should integrate over all connections compatible with asymptotically AdS boundary conditions,
\be\label{Bdefs} G(z_1,j_1; z_2, j_2; \ldots z_n , j_n) = \int\! DA_\mu e^{-S_{CS}(A)}\langle S| \prod_{i=1}^n W_{j_i}[z_i,z_b] |j_i j_i\rangle~. \ee
Rather than performing the explicit path integral we can follow \cite{Fitzpatrick:2016mtp} and take the point of view that the effect is simply to produce correlation functions of the stress tensor appearing in (\ref{adef}).  That is, we expand the path ordered exponentials in powers of $T(z)$, and then replace a string of $T(z)$ operators by the corresponding vacuum correlator, recalling that these are uniquely fixed by Virasoro symmetry.  At a formal level this recipe is justified \cite{Verlinde:1989ua} on the grounds that the objects it produces satisfy the Virasoro Ward identities, and some explicit checks of the $1/c$ expansions applied to four-point blocks were carried out in \cite{Fitzpatrick:2016mtp}.

We focus here on a two-point function since our goal is to compute scaling dimensions.  To get a nonzero result the two representations appearing in (\ref{Bdef}) should be conjugates of each other, in order that their product contain a singlet.   We then simplify by using the freedom to choose $z_b$ to place $z_b$ coincident with one of our operator insertions.  The result is that the two-point function is
\be G_j(z_1,z_2) = \langle j,-j|W_j[z_1,z_2]|j j\rangle ~.\ee
As already mentioned, we are taking $j$ to be a non-negative integer, so that we have a finite dimensional representation with states $|jm\rangle$,  $m=-j, -j+1 , \ldots j$, but this is essentially just for notational convenience.  Using the prescription of \cite{Fitzpatrick:2016mtp}, the same functional  $j$ dependence arises order by order in perturbation theory for the infinite dimensional representations.   More explicitly, we have the following
\be  G_j(z_1,z_2) = \langle j,-j| Pe^{\int_{z_1}^{z_2} a(y)dy}|jj\rangle = \sum_{n=0}^\infty \int_{z_1}^{z_2}\! dy_n \int_{z_1}^{y_n}\! dy_{n-1} \ldots \int_{z_1}^{y_2} \! dy_1 \langle j,-j|a(y_n) \ldots a(y_1)|jj\rangle~, \ee
with $a$ given in (\ref{adef}) and where each string of stress tensors is replaced by its vacuum correlator.

If the CFT operator has a definite scaling dimension the result should take the form
\be G_j(z_1,z_2) = C z_{21}^{-2h(j,c) }~,\quad z_{ij}=z_i-z_j.\ee
In the $1/c$ expansion we write
\be h(j,c) = \sum_{n=0}^\infty {h_n(j)\over c^n }~, \ee
so that
\be\label{Gjexp} G_j(z_1,z_2)  = C z_{21}^{-2h_0(j)} \left(1 - {2h_1(j)\over c} \ln z_{21} -{2h_2(j)\over c^2}\ln z_{21} +{2h_1(j)^2 \over c^2}(\ln z_{21})^2 + \ldots \right)~.\ee
The overall constant $C$ will itself have a $1/c$ expansion.     Based on our CFT discussion, we expect the results,
\be\label{hcoef}  h_0(j) = -j~,\quad h_1(j) = -6j(j+1)~,\quad h_2(j)= -78j(j+1). \ee
Our explicit computation of $G_j(z_1,z_2)$ will encounter UV divergences due to the collision of stress tensor insertions on the Wilson line.  In the analogous computation of four-point conformal blocks in \cite{Fitzpatrick:2016mtp} a normal ordering prescription was adopted such that there were no contractions between any pair of stress tensors on the same Wilson line.  That is of course not an option here, since we just have a single Wilson line and the entire result comes from such contractions.

\section{Computation of the two-point function}

\subsection{Expansion in $T(z)$}

We now perform a simple transformation so that we can expand the Wilson line in powers of $T(z)$ rather than $a(z)$.  Starting from
\be
W[z_1,z_2] = Pe^{\int_{z_1}^{z_2} \! \big( L_1+{6\over c}T(y)L_{-1}\big)dy} \ee
we define $V[z_1,z_2] = e^{-L_1 z_{21}} W[z_1,z_2] $ which obeys
\bea {d\over dz_2}  V[z_1,z_2]  &=& e^{-L_1 z_{21}} {6\over c} T(z_2)L_{-1} e^{L_1 z_{21}} V[z_1,z_2] \\ & = &  {6\over c} (L_{-1}-2z_{21}L_0 +z_{21}^2 L_1 )T(z_2) V[z_1,z_2]~.  \eea
Solving this by a path ordered exponential then yields
\be\label{Wsimp} W[z_1,z_2] = e^{L_1 z_{21} } P e^{{6\over c}\int_{z_1}^{z_2} (L_{-1}-2(y-z_1)L_0 +(y-z_1)^2 L_1 )T(y) dy}~.\ee
To implement the $1/c$ expansion we now just need to expand the second exponential factor.  To streamline our expressions we now set
\be z_2=z~,\quad z_1 =0 \ee
so that $z_{21}=z$.

\subsection{ Order $c^0$}

At leading order we have simply
\be
G^{(0)}(z) = \langle j,-j| e^{L_1 z}|jj\rangle   \sim z^{2j}~,
\ee
so that $h_0(j) =-j$ as expected.

\subsection{ Order $1/c$}

Since $\langle 0 |T(z)|0\rangle =0$ the first nontrivial correction comes from expanding the second exponential factor in (\ref{Wsimp}) to second order, yielding
\be\label{G1} G^{(1)}(z) = {6^2 \over c^2} \int_0^z\! dy_1 \int_0^{y_1} \! dy_2  \langle j,-j| e^{L_1 z} (L_{-1}-2y_1L_0 +y_1^2 L_1 )(L_{-1}-2y_2L_0 +y_2^2 L_1 )|jj\rangle \langle T(y_1)T(y_2)\rangle \ee
The SL(2,R) matrix element is easily computed by the following strategy, which extends to more complicated higher order cases.  Use the commutation relations to put the generators in the normal order $(L_1)^{n_1} (L_0)^{n_0} (L_{-1})^{n_{-1}}$.    Using $L_{-1}|jj\rangle =0$ and $L_0 |jj\rangle = j|jj\rangle$ we are left with only $L_1$ insertions, and only the power $(L_1)^{2j}$ has a nonzero matrix element.  This gives
\bea   && \langle j,-j| e^{L_1 z} (L_{-1}-2y_1L_0 +y_1^2 L_1 )(L_{-1}-2y_2L_0 +y_2^2 L_1 )|jj\rangle \\
&& \quad =    \langle j, -j|e^{L_1 z}|jj\rangle   {2jy_2(z-y_1)\big(2jy_1(z-y_2)-y_2(z-y_1)\big)  \over z^2}  \eea
As for the stress tensor correlator, we have the usual expression
\be \langle T(y_1)T(y_2)\rangle = {c/2\over (y_1-y_2)^4}~. \ee
Note that $c$ is the full central charge; this is why the self-energy diagrams of figure \ref{fig:fb}) is implicitly included.
The integral in (\ref{G1}) diverges when $y_2 \rt y_1$ and needs to be regulated.  Our strategy will be as follows.  In general, stress tensor correlators will be built out of products of factors of the form $1/(y_i - y_j)^2$, and we regulate these by making the replacement
\be\label{rglt}  {1\over (y_i -y_j)^2 }  \rt  {1\over (y_i -y_j)^2 +\eps^2  }~, \ee
so in particular we now take
\be \langle T(y_1)T(y_2)\rangle = {c/2\over \big((y_1-y_2)^2+\eps^2\big)^2}~. \ee
One way to motivate this is to express the stress tensor in terms of $c$ free bosons, $T(z) = \sum_i \p \phi_i(z) \p \phi_i(z)$.  Stress tensor correlators are then obtained by Wick's theorem.  If we regulate the basic two-point function as $\langle \p \phi(z) \p \phi(0)\rangle  = 1/(z^2+\eps^2)$ then we recover the above procedure.   The advantage of this regulator is that it is computationally tractable.  On the other hand, introducing a nonzero $\eps$ of course breaks conformal invariance, and it is not immediately obvious how to subtract divergences such that conformal invariance is recovered as $\eps \rt 0$.

We now compute
\bea  G^{(1)}(z)  & = &  \langle j, -j|e^{L_1 z}|j j\rangle { 36 j \over c}   \int_0^z\! dy_1 \int_0^{y_1}\!dy_2 {y_2(z-y_1)\big(2jy_1(z-y_2)-y_2(z-y_1)\big) \over z^2 \big((y_1-y_2)^2+\eps^2\big)^2 }  \nonumber \\
& = & \langle j, -j|e^{L_1 z}|j j\rangle { 36 j \over c}  \Bigg[  {(2j-1)\pi z^3 \over 120 \eps^3}+{z^2\over 12 \eps^2}-{(j+1)\pi z \over 12 \eps} +{j+1 \over 3} \ln {z\over \eps} +{2j-1\over 18} +O(\eps)  \Bigg]  \nonumber \\
\eea
We now perform a ``minimal subtraction" and simply remove the divergent terms and then set $\eps=0$, even though there is no clear relation at this stage to adding local counterterms to an underlying action. This gives
\be\label{G1a} G^{(1)}(z) =\langle j, -j|e^{L_1 z}|j j\rangle \Big[ {2j(2j-1) \over c} + {12j(j+1) \over c} \ln z \Big]~.\ee
Combining this with the order $c^0$ contribution, we have
\be  G^{(0)}(z)+  G^{(1)} = C z^{2j} \big[ 1 +   {12j(j+1) \over c} \ln z + O({1\over c^2})\big]~, \ee
from which we read off $h_1(j) = -6j(j+1)$ in perfect agreement with (\ref{hcoef}).

\subsection{ Order $1/c^2$}

\begin{figure}[h]
  \flushleft
  \begin{minipage}[b]{0.68\textwidth}
    \centering\includegraphics[width=1.4\textwidth]{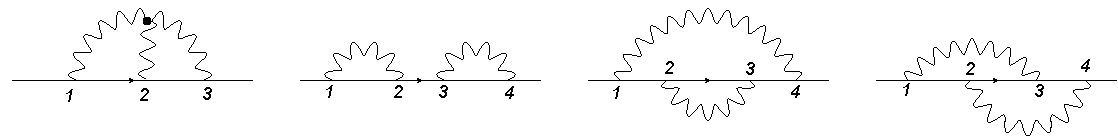}
  \end{minipage}
  \caption{ Diagrams contributing at order $1/c^2$, with stress tensor insertions on the Wilson line as indicated. }
  \label{fig:f1}
\end{figure}

At this order there are four contributing diagrams.  One diagram comes from expanding the exponential in (\ref{Wsimp}) to third order and using $\langle T(y_1)T(y_2)T(y_3)\rangle \sim c$.   However we can also expand (\ref{Wsimp}) to fourth order and use the fact that $\langle T(y_1)T(y_2)T(y_3)T(y_4)\rangle $ has order $c^2$ contributions, which can be thought of as the disconnected diagrams.   There are three such disconnected diagrams.      The four contributing diagrams are shown in figure \ref{fig:f1}.

We  evaluated these four diagrams using the approach described in the   appendix.  As in the above, we renormalize by dropping divergent terms.

\subsubsection{ $G^{(2)}_{123}(z)$}

Here we use the regulated three-point function
\be
\langle T(y_1) T(y_2)T(y_3)\rangle = {c\over [(y_1-y_2)^2+\eps^2][(y_2-y_3)^2+\eps^2][(y_3-y_1)^2+\eps^2] }  \ee
The result is
\be { G^{(2)}_{123}(z)\over \langle j, -j|e^{L_1 z}|j j\rangle} =  \Big[  -{168 j(j+1)-144j^3\over c^2} \ln z -{144 j(j+1)\over c^2} (\ln z)^2  \Big]~.\ee

\subsubsection{ $G^{(2)}_{12;34}(z)$ }

We use
\be
\langle T(y_1) T(y_2)T(y_3)T(y_4)\rangle\big|_{12;34} =  {c^2/4  \over [(y_1-y_2)^2+\eps^2]^2[(y_3-y_4)^2+\eps^2]^2} ~, \ee
which yields
\be {G^{(2)}_{12;34}(z)\over \langle j, -j|e^{L_1 z}|j j\rangle} =  \Big[  {1\over c^2}\left( {72\over 5}j-264j^2 +{384\over 5}j^3+{1776\over 5}j^4\right) \ln z   +  {144 j^2(j+1)^2\over c^2} (\ln z)^2  \Big]~.\ee

\subsubsection{ $G^{(2)}_{14;23}(z)$ }

We use
\be
\langle T(y_1) T(y_2)T(y_3)T(y_4)\rangle\big|_{14;23} =  {c^2/4  \over [(y_1-y_4)^2+\eps^2]^2[(y_2-y_3)^2+\eps^2]^2} ~, \ee
which yields
\be {G^{(2)}_{14;23}(z)\over \langle j, -j|e^{L_1 z}|j j\rangle} =  \Big[  {1\over c^2}\left( -{324\over 5}j -{492\over 5}j^2+{1824\over 5}j^3 +{1992\over 5}j^4\right) \ln z   +  {72 j^2(j+1)^2\over c^2} (\ln z)^2  \Big]~.\ee

\subsubsection{ $G^{(2)}_{13;24}(z)$ }

We use
\be
\langle T(y_1) T(y_2)T(y_3)T(y_4)\rangle\big|_{13;24} =  {c^2/4  \over [(y_1-y_3)^2+\eps^2]^2[(y_2-y_4)^2+\eps^2]^2} ~, \ee
which yields
\be {G^{(2)}_{13;24}(z)\over \langle j, -j|e^{L_1 z}|j j\rangle} =  \Big[  {1\over c^2}\left({396\over 5}j +{1908\over 5}j^2-{2736\over 5}j^3 -{3528\over 5}j^4\right) \ln z   +  {1\over c^2}\left(144j -288j^3-144j^4\right) (\ln z)^2  \Big]~.\ee

\subsubsection{Complete result at order $1/c^2$}

We now combine all of our results for the complete correlator up to this order.   The result is
\bea G(z) &=& G^{(0)}(z)+G^{(1)}(z)+G^{(2)}(z) + \ldots \nonumber \\
&=& C z^{2j} \Bigg[1 +{12 j(j+1)\over c}\ln z +{24 (3j-29)j(j+1)\over 5c^2}\ln z +{72 j^2(j+1)^2 \over c^2} (\ln z)^2 +O\left({1\over c^3}\right)   \Bigg]  \nonumber \\ \eea
Note that the $2j(2j-1)/c$ term in (\ref{G1a}) contributed to this, since we have set the leading term in $[\ldots]$ to be $1$ by absorbing the overall constant factor into $C$.

Comparing with expectations, we see that the $(\ln z)^2$ term is in agreement with (\ref{Gjexp}), so that the result to this order takes the form of a single power of $z$.   This is quite nontrivial from the diagrammatic point of view, as there are $(\ln z)^2$ contributions from all four of the $1/c^2$ diagrams which must all combine together to give the correct coefficient.  On the other hand, the ${1\over c^2} \ln z$ term does not have the expected  coefficient  $-2h_2(j)/c^2 = 156 j(j+1)/c^2$.

We now make a few comments about this result.   A feature that emerges at order $1/c^2$ but which is absent at order $1/c$ is the appearance of divergent terms of the form ${1\over c^2 \eps^n}\ln z$.  If we take the general point of view that when removing a divergence we can also subtract a finite term with the same $z$ dependence, then this renders the coefficient of the ${1\over c^2}\ln z$ term ambiguous.  By contrast, the absence of divergences of the form ${1\over c \eps^n}\ln z$ and ${1\over c^2 \eps^n}(\ln z)^2$ suggests that the coefficients of the terms ${1\over c}\ln z$ and ${1\over c^2 }(\ln z)^2$ are unambiguous, and indeed these coefficients precisely match expectations.   Of course, what this emphasizes is the need for a more systematic renormalization approach.  On the other hand, we again note the fact that our result to this order takes the form of a single power law in $z$, suggesting that conformal invariance is being respected by our procedure.

\section{Discussion}

We have computed the expectation value of an open Wilson line to order $1/c^2$.  From this result we read off the scaling dimension of the corresponding primary operator and compared it to expectations from CFT considerations.  This revealed partial agreement with CFT predictions as well as some unresolved issues.  The order $1/c$ result was as expected, and furthermore we found that at order $1/c^2$ the result takes the form of a single power law, as dictated by conformal invariance.  On the other hand, the order $1/c^2$ contribution to the scaling dimension is at odds with our expectations.  More accurately, the result is ambiguous within the framework of our computation, as the desired coefficient of a ${1\over c^2}\ln z$ term is ``corrupted" by the presence of ${1\over \eps^n c^2}\ln z$ UV divergences requiring renormalization.  This clearly points to the need for a more principled renormalization scheme.

There are of course other ways to regulate the stress tensor correlators. For example, instead of making the replacement in (\ref{rglt}) we can implement a simple version of dimensional regularization.  In particular, we can  replace the exponent 2 in the denominator with $(2-\epsilon)$, taking $\epsilon$ to be sufficiently positive so that the integrals converge, and then  analytically continue the result to $\epsilon$ near $0$.   After a minimal subtraction of pole terms, the $1/c$ contribution we find is still in agreement with (\ref{hcoef}) but the $1/c^2$ contribution is not.

It is worth contrasting what we have found here with what one encounters in the computation of closed Wilson loops in ordinary Chern-Simons theory, which yield topological invariants \cite{Witten:1988hf}.   The leading order contribution comes from a gluon exchanged between two points on the Wilson loop.  This leads to an integral which is UV finite, but the result is not a topological invariant.  To rectify this one needs to introduce a ``framing", corresponding to displacing the worldines on which the two gluons are inserted.  The result is a topological invariant that depends on the choice of framing \cite{Witten:1988hf,Guadagnini:1989am}.  

Our primary operators are labelled by an SL(2,R) spin $j$, which from the CFT side comes from constructing Virasoro representations by applying constraints to SL(2,R) current algebra representations. An SL(2,R) spin-$j$ also naturally appears in the bulk, via the formulation of gravity in terms of SL(2,R) Chern-Simons theory, and it therefore seems meaningful to compare scaling dimensions in the two descriptions as a function of $j$ and the central charge $c$.  On the other hand, strictly from the Virasoro point of view, $j$ is simply a label, so one might wonder if there is perhaps some $c$ dependent relation between the $j$ labels in the two descriptions.  To address this we note that degenerate representations correspond to $2j$ being a positive integer, which precludes such a $c$ dependent relation for such representations.  This is to say that we certainly expect to be able to meaningfully compare the scaling dimensions of degenerate representations on the two sides as a function of $c$.   Of course, these scaling dimensions are entirely fixed by Virasoro representation theory, but we do not want to use this, as the entire point here is to develop computational rules in the bulk that will apply more generally.

We have tried to extract scaling dimensions from two-point functions, but another approach is to adopt  canonical quantization \cite{Elitzur:1989nr}. In particular, we can consider a single particle, associated to a spin-$j$ representation of SL(2,R), coupled in a gauge invariant fashion to SL(2,R) Chern-Simons gauge fields.  One should be able to realize the Virasoro generators on this  Hilbert space, and demanding that the algebra is realized consistently may resolve the ambiguities associated with renormalizing UV divergences.   We hope to report on this in the near future.

\section*{Acknowledgements}

We thank Thomas Dumitrescu for useful discussions.   P.K. is supported in part by NSF grant PHY-1313986.

\appendix

\section{Evaluation of integrals}

We encounter nested integrals of the form
\be I(z) = \int_0^z \! dy_1 \int_0^{y_1} \! dy_2 \ldots \int_0^{y_{n-1}}\! dy_n  { P(z,y_i) \over \prod_{i<j} [(y_i-y_j)^2+\eps^2]^{n_{ij}}}~,\ee
where  $P(z,y_i)$ is a polynomial and $n_{ij}$ are non-negative integers.

We first rewrite this in terms of unconstrained integrals by introducing step functions,
\be \int_0^z \! dy_1 \int_0^{y_1} \! dy_2 \ldots \int_0^{y_{n-1}}\! dy_n \rt  \int\! d^n y ~ \theta(z-y_1)\theta(y_1-y_2) \ldots \theta(y_{n-1}-y_n)\theta(y_n)~,\ee
and use the Fourier representation
\be \theta(y) = \int\! {dp\over 2\pi i } {e^{ipy}\over p- i\delta }~,\quad \delta >0~. \ee
We also write  the denominator factors in momentum space using
\be {1\over y^2+\eps^2} = {1\over 2\eps} \int_{-\infty}^\infty \! dk e^{iky-|k|\eps}~.\ee
The $y$-integrals can then be carried out, yielding $n$ delta functions involving $p$ and $k$.  These delta functions soak up all but one of the $p$ integrals, and the remaining $p$ integral can be done by computing residues.   This leaves some $k$-integrals, where the integrand is a sum of terms taking the form of exponentials time rational functions.  Some of the denominator factors can be removed by differentiating with respect to $z$, and the other by using relations like
\be\label{kid} {1\over k_1 -k_2} =-{i\over 2}  \int_{-\infty}^\infty \! du ~{\rm sgn}(u) e^{i(k_1-k_2)u}~. \ee
The $k$ integrals are then carried out, followed by the $u$ integrals.  The result is then expanded for small $\eps$, and we finally integrate to undo the earlier $z$ differentiation.  Due to the last step, this procedure will only determine the result up to a polynomial in $z$. However, if desired, this polynomial can easily be determined by directly studing the small $z$ expansion of the original integral.

We present a representative example to make the procedure  concrete,
\be  I_3(z) = \int_0^z\! dy_1 \int_0^{y_1}\! dy_2 \int_0^{y_2}\! dy_3 {1\over (y_1-y_2)^2+\eps^2}{1\over (y_1-y_3)^2+\eps^2}{1\over (y_2-y_3)^2+\eps^2}~. \ee
Proceeding as above, we have

\bea I_3(z)&=& {1\over 8 \eps^3} \int\! {d^4p d^3k \over (2\pi i)^4} {e^{-|k_1|\eps-|k_2|\eps-|k_3|\eps}\over (p_1-i\delta)(p_2-i\delta)(p_3-i\delta)(p_4-i\delta)}    \nonumber \\
&& \quad \times  \int\! d^3y e^{ip_1(z-y_1)+ip_2y_{12}+ip_3y_{23}+ip_4y_3+ik_1 y_{12}+ik_2 y_{23}+ik_3 y_{31}}  \nonumber \\
& =&  {i\over 8\eps^3} \int\! {dp_1 d^3k \over 2\pi i} {e^{-|k_1|\eps-|k_2|\eps-|k_3\eps}\over (p_1-i\delta)^2(p_1-k_1+k_3-i\delta)(p_1-k_2+k_3-i\delta)}e^{ip_1z}  \nonumber \\
& =&  {i\over 8 \eps^3} \int\! d^3k  e^{-|k_1|\eps-|k_2|\eps-|k_3|\eps}\Bigg[ {e^{i(k_1-k_3)z}\over (k_1-k_3)^2(k_1-k_2)}-{e^{i(k_2-k_3)z}\over (k_2-k_3)^2(k_1-k_2)}  \nonumber \\
 &&\quad \quad \quad \quad \quad \quad \quad \quad \quad \quad \quad +{ (k_1k_2+k_3^2-k_1k_3-k_2k_3)z+k_1+k_2-2k_3 \over (k_1-k_3)^2(k_2-k_3)^2} \Bigg]   \nonumber \\\eea
In getting to the final expression we performed the $p_1$ integral by residues, but discarded the contribution from the pole at $p_1=i\delta$, since this will only contribute a degree 1 polynomial in $z$ that will anyway be killed by the derivatives that we will apply in the next step.  On the other hand, convergence of the $k$ integrals in the above undifferentiated expression does require the presence of this polynomial part, as it is needed to render the integrand finite at the locations where the denominator factors vanish.

We now differentiate twice to get
\bea {\p^2 I_3 \over \p z^2} &=& - {i\over 8\eps^3} \int\! d^3k e^{-|k_1|\eps-|k_2|\eps-|k_3|\eps}e^{-ik_3z} {e^{ik_1z}-e^{ik_2z}\over k_1-k_2} \nonumber \\
& = & - {i\over 4\eps^2} {1\over z^2+\eps^2} \int\!  d^2k  e^{-|k_1|\eps-|k_2|\eps} {e^{ik_1z}-e^{ik_2z}\over k_1-k_2}  \eea
Using (\ref{kid}) gives
\bea  {\p^2 I_3 \over \p z^2} &=&  - {1\over 8\eps^2} {1\over z^2+\eps^2}\int_{-\infty}^\infty \! du~{\rm sgn}(u)  \int\!  d^2k  e^{i(k_1-k_2)u}e^{-|k_1|\eps-|k_2|\eps} \left(e^{ik_1z}-e^{ik_2z} \right) \nonumber \\
& = & {2z\over z^2 +\eps^2}\int_{-\infty}^\infty \! du~{\rm sgn}(u)  {u\over (u^2+\eps^2)[(u+z)^2+\eps^2][(u-z)^2+\eps^2]}   \nonumber \\
 & = & 2  { \tan^{-1}\left({z\over \eps}\right) + { \eps\over z} \ln\left(1+{z^2\over \eps^2}\right) \over  \eps (z^2 +\eps^2) (z^2+4\eps^2)}   \nonumber \\
&=&  {\pi \over \eps z^4} + {4\over z^5}\ln {z\over \eps} -{2\over z^5} + O(\eps)   \eea
and so we arrive at
\be  I_3(z) = {\pi \over 6 \eps z^2}+{1\over 3z^3}\ln {z\over \eps} +{1\over 36 z^3} + O(\eps)~, \ee
where we fixed the integration constants by examining the original integral.

All of our integrals can be worked out this way.  This somewhat circuitous procedure has the advantage that it can easily be automated.

\bibliographystyle{ssg}
\bibliography{biblio}

\begingroup\raggedright\begin{thebibliography}{10}

\bibitem{Bershadsky:1989mf}
M.~Bershadsky and H.~Ooguri, ``{Hidden SL(n) Symmetry in Conformal Field
  Theories},'' {\em Commun. Math. Phys.} {\bf 126} (1989) 49.

\bibitem{Brown:1986nw}
J.~D. Brown and M.~Henneaux, ``{Central Charges in the Canonical Realization of
  Asymptotic Symmetries: An Example from Three-Dimensional Gravity},'' {\em
  Commun. Math. Phys.} {\bf 104} (1986) 207--226.

\bibitem{Banados:1998gg}
M.~Banados, ``{Three-dimensional quantum geometry and black holes},''
  \href{http://xxx.lanl.gov/abs/hep-th/9901148}{{\tt hep-th/9901148}}. [AIP
  Conf. Proc.484,147(1999)].

\bibitem{Witten:1988hc}
E.~Witten, ``{(2+1)-Dimensional Gravity as an Exactly Soluble System},'' {\em
  Nucl. Phys.} {\bf B311} (1988) 46.

\bibitem{Achucarro:1987vz}
A.~Achucarro and P.~K. Townsend, ``{A Chern-Simons Action for Three-Dimensional
  anti-De Sitter Supergravity Theories},'' {\em Phys. Lett.} {\bf B180} (1986)
  89.

\bibitem{Ammon:2013hba}
M.~Ammon, A.~Castro, and N.~Iqbal, ``{Wilson Lines and Entanglement Entropy in
  Higher Spin Gravity},'' {\em JHEP} {\bf 10} (2013) 110,
  \href{http://xxx.lanl.gov/abs/1306.4338}{{\tt 1306.4338}}.

\bibitem{deBoer:2013vca}
J.~de~Boer and J.~I. Jottar, ``{Entanglement Entropy and Higher Spin Holography
  in AdS$_3$},'' {\em JHEP} {\bf 04} (2014) 089,
  \href{http://xxx.lanl.gov/abs/1306.4347}{{\tt 1306.4347}}.

\bibitem{Hartman:2013mia}
T.~Hartman, ``{Entanglement Entropy at Large Central Charge},''
  \href{http://xxx.lanl.gov/abs/1303.6955}{{\tt 1303.6955}}.

\bibitem{Fitzpatrick:2014vua}
A.~L. Fitzpatrick, J.~Kaplan, and M.~T. Walters, ``{Universality of
  Long-Distance AdS Physics from the CFT Bootstrap},'' {\em JHEP} {\bf 08}
  (2014) 145, \href{http://xxx.lanl.gov/abs/1403.6829}{{\tt 1403.6829}}.

\bibitem{deBoer:2014sna}
J.~de~Boer, A.~Castro, E.~Hijano, J.~I. Jottar, and P.~Kraus, ``{Higher spin
  entanglement and $ {\mathcal{W}}_{\mathrm{N}} $ conformal blocks},'' {\em
  JHEP} {\bf 07} (2015) 168, \href{http://xxx.lanl.gov/abs/1412.7520}{{\tt
  1412.7520}}.

\bibitem{Hijano:2015rla}
E.~Hijano, P.~Kraus, and R.~Snively, ``{Worldline approach to semi-classical
  conformal blocks},'' {\em JHEP} {\bf 07} (2015) 131,
  \href{http://xxx.lanl.gov/abs/1501.02260}{{\tt 1501.02260}}.

\bibitem{Fitzpatrick:2015zha}
A.~L. Fitzpatrick, J.~Kaplan, and M.~T. Walters, ``{Virasoro Conformal Blocks
  and Thermality from Classical Background Fields},''
  \href{http://xxx.lanl.gov/abs/1501.05315}{{\tt 1501.05315}}.

\bibitem{Alkalaev:2015wia}
K.~B. Alkalaev and V.~A. Belavin, ``{Classical conformal blocks via AdS/CFT
  correspondence},'' {\em JHEP} {\bf 08} (2015) 049,
  \href{http://xxx.lanl.gov/abs/1504.05943}{{\tt 1504.05943}}.

\bibitem{Hijano:2015qja}
E.~Hijano, P.~Kraus, E.~Perlmutter, and R.~Snively, ``{Semiclassical Virasoro
  Blocks from AdS$_3$ Gravity},''
  \href{http://xxx.lanl.gov/abs/1508.04987}{{\tt 1508.04987}}.

\bibitem{Fitzpatrick:2015dlt}
A.~L. Fitzpatrick and J.~Kaplan, ``{Conformal Blocks Beyond the Semi-Classical
  Limit},'' \href{http://xxx.lanl.gov/abs/1512.03052}{{\tt 1512.03052}}.

\bibitem{Bhatta:2016hpz}
A.~Bhatta, P.~Raman, and N.~V. Suryanarayana, ``{Holographic Conformal Partial
  Waves as Gravitational Open Wilson Networks},''
  \href{http://xxx.lanl.gov/abs/1602.02962}{{\tt 1602.02962}}.

\bibitem{Hijano:2015zsa}
E.~Hijano, P.~Kraus, E.~Perlmutter, and R.~Snively, ``{Witten Diagrams
  Revisited: The AdS Geometry of Conformal Blocks},'' {\em JHEP} {\bf 01}
  (2016) 146, \href{http://xxx.lanl.gov/abs/1508.00501}{{\tt 1508.00501}}.

\bibitem{Alkalaev:2015lca}
K.~B. Alkalaev and V.~A. Belavin, ``{Monodromic vs geodesic computation of
  Virasoro classical conformal blocks},'' {\em Nucl. Phys.} {\bf B904} (2016)
  367--385, \href{http://xxx.lanl.gov/abs/1510.06685}{{\tt 1510.06685}}.

\bibitem{Alkalaev:2016ptm}
K.~B. Alkalaev and V.~A. Belavin, ``{Holographic interpretation of 1-point
  toroidal block in the semiclassical limit},'' {\em JHEP} {\bf 06} (2016) 183,
  \href{http://xxx.lanl.gov/abs/1603.08440}{{\tt 1603.08440}}.

\bibitem{Fitzpatrick:2016ive}
A.~L. Fitzpatrick, J.~Kaplan, D.~Li, and J.~Wang, ``{On information loss in
  AdS$_{3}$/CFT$_{2}$},'' {\em JHEP} {\bf 05} (2016) 109,
  \href{http://xxx.lanl.gov/abs/1603.08925}{{\tt 1603.08925}}.

\bibitem{Czech:2016xec}
B.~Czech, L.~Lamprou, S.~McCandlish, B.~Mosk, and J.~Sully, ``{A Stereoscopic
  Look into the Bulk},'' {\em JHEP} {\bf 07} (2016) 129,
  \href{http://xxx.lanl.gov/abs/1604.03110}{{\tt 1604.03110}}.

\bibitem{daCunha:2016crm}
B.~Carneiro~da Cunha and M.~Guica, ``{Exploring the BTZ bulk with boundary
  conformal blocks},'' \href{http://xxx.lanl.gov/abs/1604.07383}{{\tt
  1604.07383}}.

\bibitem{Chen:2016dfb}
B.~Chen, J.-q. Wu, and J.-j. Zhang, ``{Holographic Description of 2D Conformal
  Block in Semi-classical Limit},'' {\em JHEP} {\bf 10} (2016) 110,
  \href{http://xxx.lanl.gov/abs/1609.00801}{{\tt 1609.00801}}.

\bibitem{Alkalaev:2016rjl}
K.~B. Alkalaev, ``{Many-point classical conformal blocks and geodesic networks
  on the hyperbolic plane},'' {\em JHEP} {\bf 12} (2016) 070,
  \href{http://xxx.lanl.gov/abs/1610.06717}{{\tt 1610.06717}}.

\bibitem{Guica:2016pid}
M.~Guica, ``{Bulk fields from the boundary OPE},''
  \href{http://xxx.lanl.gov/abs/1610.08952}{{\tt 1610.08952}}.

\bibitem{Hegde:2015dqh}
A.~Hegde, P.~Kraus, and E.~Perlmutter, ``{General Results for Higher Spin
  Wilson Lines and Entanglement in Vasiliev Theory},'' {\em JHEP} {\bf 01}
  (2016) 176, \href{http://xxx.lanl.gov/abs/1511.05555}{{\tt 1511.05555}}.

\bibitem{Besken:2016ooo}
M.~Besken, A.~Hegde, E.~Hijano, and P.~Kraus, ``{Holographic conformal blocks
  from interacting Wilson lines},'' {\em JHEP} {\bf 08} (2016) 099,
  \href{http://xxx.lanl.gov/abs/1603.07317}{{\tt 1603.07317}}.

\bibitem{Fitzpatrick:2016mtp}
A.~L. Fitzpatrick, J.~Kaplan, D.~Li, and J.~Wang, ``{Exact Virasoro Blocks from
  Wilson Lines and Background-Independent Operators},''
  \href{http://xxx.lanl.gov/abs/1612.06385}{{\tt 1612.06385}}.

\bibitem{Verlinde:1989ua}
H.~L. Verlinde, ``{Conformal Field Theory, 2-$D$ Quantum Gravity and
  Quantization of Teichmuller Space},'' {\em Nucl. Phys.} {\bf B337} (1990)
  652--680.

\bibitem{Witten:1988hf}
E.~Witten, ``{Quantum Field Theory and the Jones Polynomial},'' {\em Commun.
  Math. Phys.} {\bf 121} (1989) 351--399.

\bibitem{Saida:1999ec}
H.~Saida and J.~Soda, ``{Statistical entropy of BTZ black hole in higher
  curvature gravity},'' {\em Phys. Lett.} {\bf B471} (2000) 358--366,
  \href{http://xxx.lanl.gov/abs/gr-qc/9909061}{{\tt gr-qc/9909061}}.

\bibitem{Kraus:2005vz}
P.~Kraus and F.~Larsen, ``{Microscopic black hole entropy in theories with
  higher derivatives},'' {\em JHEP} {\bf 09} (2005) 034,
  \href{http://xxx.lanl.gov/abs/hep-th/0506176}{{\tt hep-th/0506176}}.

\bibitem{Deser:1983tn}
S.~Deser, R.~Jackiw, and G.~'t~Hooft, ``{Three-Dimensional Einstein Gravity:
  Dynamics of Flat Space},'' {\em Annals Phys.} {\bf 152} (1984) 220.

\bibitem{Raeymaekers:2014kea}
J.~Raeymaekers, ``{Quantization of conical spaces in 3D gravity},'' {\em JHEP}
  {\bf 03} (2015) 060, \href{http://xxx.lanl.gov/abs/1412.0278}{{\tt
  1412.0278}}.

\bibitem{Fitzpatrick:2016mjq}
A.~Liam~Fitzpatrick and J.~Kaplan, ``{On the Late-Time Behavior of Virasoro
  Blocks and a Classification of Semiclassical Saddles},''
  \href{http://xxx.lanl.gov/abs/1609.07153}{{\tt 1609.07153}}.

\bibitem{Guadagnini:1989am}
E.~Guadagnini, M.~Martellini, and M.~Mintchev, ``{Wilson Lines in Chern-Simons
  Theory and Link Invariants},'' {\em Nucl. Phys.} {\bf B330} (1990) 575--607.

\bibitem{Elitzur:1989nr}
S.~Elitzur, G.~W. Moore, A.~Schwimmer, and N.~Seiberg, ``{Remarks on the
  Canonical Quantization of the Chern-Simons-Witten Theory},'' {\em Nucl.
  Phys.} {\bf B326} (1989) 108--134.

\end{thebibliography}\endgroup

\end{document}